\documentclass[review,3p,twocolumn]{elsarticle}

\usepackage[hidelinks]{hyperref}

\usepackage{booktabs}       
\usepackage{pgfplots}       
\usepackage{pgfplotstable}  
\usepackage{tikz}           
\usetikzlibrary{patterns}
\usepackage{subcaption}

\makeatletter
\def\ps@pprintTitle{%
   \let\@oddhead\@empty
   \let\@evenhead\@empty
   \let\@oddfoot\@empty
   \let\@evenfoot\@oddfoot
}
\makeatother

\usepackage{listings}
\definecolor{strings_color}{rgb}{0.49,0.49,0.49}
\definecolor{comments_color}{rgb}{0.49,0.49,0.49}
\definecolor{border_color}{rgb}{0.81,0.81,0.81}
\usepackage{caption}
\usepackage{chngcntr} 

\journal{Arxiv.org}

\begin{document}

\begin{frontmatter}

\title{Decoupled Strategy for Imbalanced Workloads in MapReduce Frameworks}





\author[kth]{Sergio~Rivas-Gomez}
\ead{sergiorg@kth.se}
\author[sea]{Sai~Narasimhamurthy}
\ead{sai.narasimhamurthy@seagate.com}
\author[arm]{Keeran~Brabazon}
\ead{keeran.brabazon@arm.com}
\author[arm]{Oliver~Perks}
\ead{olly.perks@arm.com}
\author[kth]{Erwin~Laure}
\ead{erwinl@kth.se}
\author[kth]{Stefano~Markidis}
\ead{markidis@kth.se}
\address[kth]{KTH Royal Institute of Technology, Stockholm 10044, Sweden}
\address[sea]{Seagate Systems UK, Havant PO9 1SA, UK}
\address[arm]{Arm, Cambridge CB1 9NJ, UK}

\begin{abstract}
In this work, we consider the integration of MPI one-sided communication and non-blocking I/O in HPC-centric MapReduce frameworks. Using a decoupled strategy, we aim to overlap the Map and Reduce phases of the algorithm by allowing processes to communicate and synchronize using solely one-sided operations. Hence, we effectively increase the performance in situations where the workload per process is unexpectedly unbalanced. Using a Word-Count implementation and a large dataset from the Purdue MapReduce Benchmarks Suite (PUMA), we demonstrate that our approach can provide up to 23\% performance improvement on average compared to a reference MapReduce implementation that uses state-of-the-art MPI collective communication and I/O.
\end{abstract}


\end{frontmatter}

\section{Introduction}
\label{1_Introduction}

During the past decade, data-intensive workloads have become an integral
part of large-scale scientific computing~\cite{sage2016whitepaper, gao2017mimir}. The emergence of machine learning and data-centric applications on HPC, has been motivated by the advances in deep learning and convolutional networks~\cite{szegedy2015going}, alongside with the appearance of programming models and tools for data mining~\cite{witten2016data}. These developments allow us to understand large datasets of unstructured information.

In this regard, MapReduce has become one of the preferred programming models to hide the complexity of process and data parallelism~\cite{dean2008mapreduce, ekanayake2008mapreduce}. The power of this paradigm resides on the definition of simple \texttt{Map()} and \texttt{Reduce()} functions, that become highly-parallel operations using complex inter-processor communication~\cite{hoefler2009towards}. For instance, this model has been successfully applied in the past for the trajectory analysis of high-performance molecular dynamics (MD) simulations~\cite{tu2008scalable}, that model important biological processes that occur on the millisecond time scale.

Despite traditional HPC clusters offering immense potential for high-performance data analytics, it has also been debated that MapReduce frameworks, such as Hadoop MapReduce~\cite{shvachko2010hadoop}, pose numerous constraints on current supercomputers due to the elevated memory / storage requirements and complex job scheduling~\cite{reed2015exascale, gao2017mimir}. Consequently, over the past few years, MPI-based implementations of MapReduce have originated with the aim of taking advantage of the high-performance network and storage subsystems of HPC clusters~\cite{plimpton2011mapreduce, guo2015fault}. These implementations integrate the highly-optimized collective communication and I/O of MPI within the different phases of the algorithm.

Notwithstanding, as the concurrency of upcoming HPC clusters is expected to increase 100--1000$\times$~\cite{dongarra2011international}, several limitations arise from the use of master-slave or the inherent coupling between the Map and Reduce phases of traditional MapReduce frameworks. Given the irregular nature of certain input datasets, these design considerations pose performance restrictions when the workload per process becomes unexpectedly unbalanced. In such cases, it has been demonstrated that the use of decentralized algorithms can provide significant performance benefits~\cite{peng2017preparing}.

In this work, we set the initial steps towards the integration of a decoupled strategy for MapReduce on HPC. In particular, we design and implement \emph{MapReduce-1S} (i.e., MapReduce ``One-Sided''), a small MapReduce implementation that uses MPI one-sided communication and non-blocking I/O~\cite{thakur1999data, gropp2014using} to overlap the Map and Reduce phases of the algorithm. Processes synchronize using solely one-sided operations. For instance, conventional \texttt{put} / \texttt{get} operations are employed to update and retrieve the key-value pairs remotely. Moreover, the distribution of the tasks during Map is also decentralized and self-managed (i.e., each process reads non-overlapping portions of the input datasets independently).

Using a Word-Count implementation and a dataset from the Purdue MapReduce Benchmarks Suite (PUMA)~\cite{ahmad2012puma}, we demonstrate that our approach provides up to 23\% performance improvement on unbalanced workloads compared to a reference MapReduce implementation that uses state-of-the-art MPI collective operations~\cite{hoefler2009towards}. On perfectly balanced workloads, however, we also observe that the opportunities for improvement are negligible. Nonetheless, \mbox{MapReduce-1S} can still provide additional advantages, such as novel fault-tolerance support with the integration of the MPI storage windows concept~\cite{rivas2017mpi}.

The contributions of this work are the following:
    
\begin{itemize}
\item We design and develop MapReduce-1S, an implementation of MapReduce based on the use of MPI one-sided communication and non-blocking IO.
\item We provide a custom MapReduce framework to support our experiments with MapReduce-1S.
\item We illustrate the performance of MapReduce-1S on Word-Count under balanced and unbalanced workloads, using strong / weak scaling evaluations.
\item We provide initial performance measurements for a fault-tolerant implementation of MapReduce-1S.
\end{itemize}

The paper is organized as follows. We present the design and implementation details of MapReduce-1S in \autoref{2_Methods}. The experimental setup and performance results on Word-Count are presented in \autoref{3_Results}. We extend the discussion of the results and provide further insights in \autoref{4_Discussion}. Related work is briefly described in \autoref{5_RelatedWork}. Lastly, \autoref{6_Conclusion} summarizes our conclusions and outlines future work.

\section{Decoupled MapReduce}
\label{2_Methods}

MapReduce emerged in the context of cloud analytics as a programming model for processing and generating large datasets~\cite{dean2008mapreduce, reed2015exascale}. The main idea behind a MapReduce job is to split a certain input dataset into independent portions or \emph{tasks}, which can then be processed in a completely parallel manner inside the \emph{Map} phase. The output from this phase is sorted during an intermediate step called \emph{Shuffle}. Finally, the ordered output is transferred to the \emph{Reduce} phase, where the data is aggregated to produce the result.

Inside a MapReduce framework, users are responsible for the implementation of the \texttt{Map()} and \texttt{Reduce()} operations. In particular, \texttt{Map()} is designed to split the input data into a collection of individual key-value pairs. Each tuple is then merged using the \texttt{Reduce()} function, producing an aggregation of all the key-value pairs with identical key. Despite its simplicity, many real-world applications can be expressed following this model, such as high-energy physics data analysis, or K-means clustering~\cite{chu2007map, ekanayake2008mapreduce}.

\begin{figure}
    \begin{center}
        \hspace{-0.81cm}
        \includegraphics[width=0.81\columnwidth]{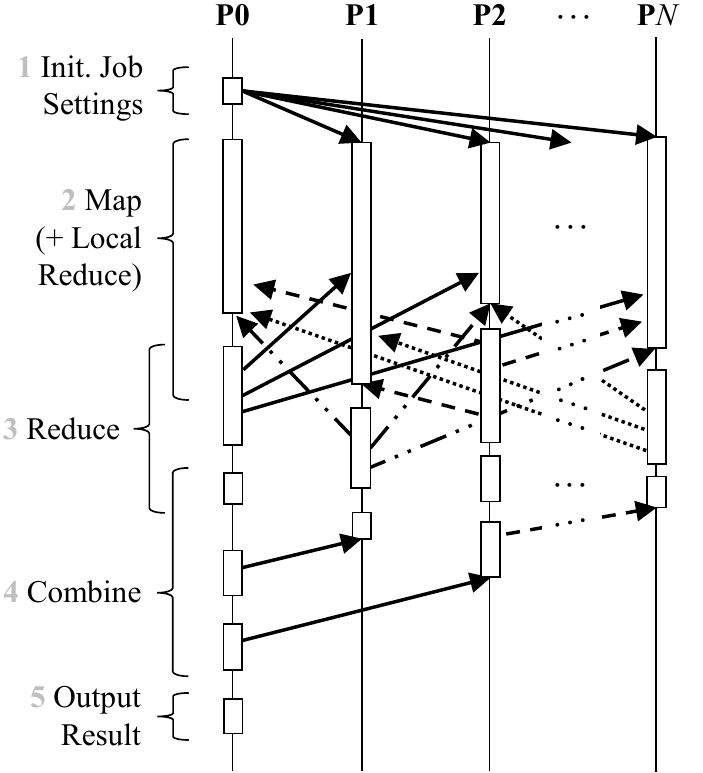}
        \caption{The use of MPI one-sided communication and non-blocking I/O enables an effective overlap of the Map and Reduce phases, specially on unbalanced workloads.}
        \label{fig:MapReduce1S}
    \end{center}
\end{figure}

In the context of HPC, MapReduce implementations are frequently integrated using state-of-the-art MPI functionality~\cite{hoefler2009towards, plimpton2011mapreduce}. For instance, tasks are commonly distributed employing a master-slave approach with scatter operations (e.g., \texttt{MPI\_Scatter}). Fixed-length, associative key-values can be used to take advantage of the heavily-optimized reduce operations of MPI (e.g., \texttt{MPI\_Reduce}). In addition, the intermediate data-shuffle can be mapped to collective \mbox{all-to-all} operations (e.g., \texttt{MPI\_Alltoall}), optimizing the communication between the different phases~\cite{guo2015fault}. When reading the input, MPI collective I/O can be used to decrease the overhead of accessing parallel file systems~\cite{thakur1999data,gao2017mimir}.

Even though these design considerations generally provide major advantages compared to cloud-based alternatives~\cite{lu2011can}, we also observe that the inherent coupling between the Map and Reduce phases may still produce workload imbalance. This is particularly the case when the input datasets feature an irregular distribution of the data. In such cases, a decoupled strategy for MapReduce frameworks can reduce the synchronization overhead among the processes by overlapping the different phases of the algorithm. We propose to solve this challenge with the integration of the MPI one-sided communication model~\cite{gropp2014using, gerstenberger2014enabling}. This model enables local / remote communication over special-buffers denoted as MPI \emph{windows}. The basic operations defined by the MPI standard to access and update an MPI window are \texttt{put} and \texttt{get}. Advanced functionality, such as atomic Compare-And-Swap (CAS), is also available. In addition, we also consider the integration of MPI non-blocking I/O~\cite{thakur1999data} to overlap computations and storage operations.

In this section, we present the design and implementation details of MapReduce-1S, a small MapReduce implementation that uses MPI one-sided communication and non-blocking I/O. We additionally describe a custom MapReduce framework utilized to support our experiments.


\subsection{Design and Implementation}

MapReduce-1S inherits the core principles of traditional MapReduce frameworks, such as Hadoop MapReduce~\cite{shvachko2010hadoop}, with subtle variations. In particular, we opt to divide the execution into four different isolated phases (\autoref{fig:MapReduce1S}):

\renewcommand{\labelenumi}{\Roman{enumi}}
\begin{enumerate}
\item \textbf{Map}. Transforms a given input into multiple key-value pairs. Each key-value is assigned to a target process and stored into a designated buffer for remote communication. The ownership is determined through a \emph{hash} function using the key.
\item \textbf{Local Reduce}. Aggregates certain key-value pairs locally, whenever possible. The aim is to decrease the overall memory footprint and network overhead~\cite{gao2017mimir}. This phase is conducted within Map.
\item \textbf{Reduce}. Aggregates all the key-value pairs found by the rest of the processes. Remote memory operations are used to retrieve the tuples. The output is an ordered collection of unique key-value pairs stored locally.
\item \textbf{Combine}. Combines the aggregated key-value pairs to generate the final result. This phase is similar to \emph{Shuffle} in traditional MapReduce frameworks, with the difference that Reduce also performs ordering. Hence, this step is considerably lighter.
\end{enumerate}

The input datasets are split into equally-sized tasks, that are later handled in parallel by each process. Instead of following a master-slave approach, we design a mechanism that enable processes to decide the next task to perform based on the rank, task size, and file offset between tasks. The input portion for the task is retrieved individually using non-blocking MPI I/O operations. Hence, while a certain task is being computed, the subsequent input is already scheduled for asynchronous retrieval.

\begin{figure}
    \begin{center}
        \includegraphics[width=1.0\columnwidth]{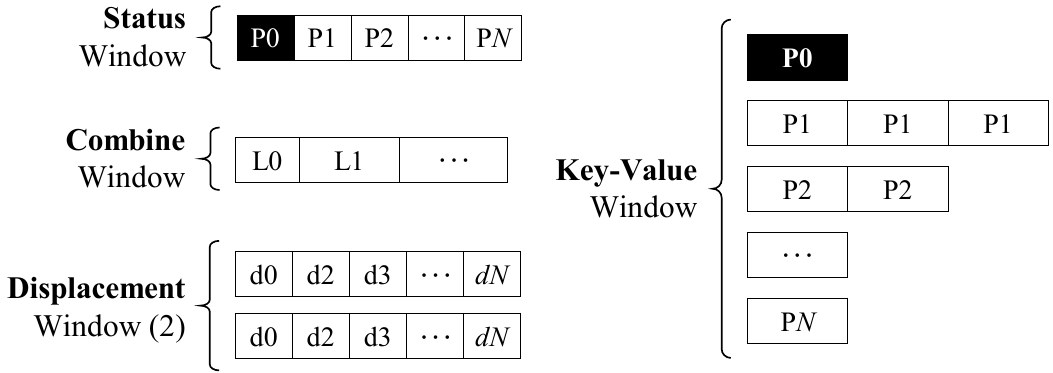}
        \vspace{-0.21cm}
        \caption{Multiple MPI windows are required to support the decoupled strategy of MapReduce-1S. This allows processes to communicate using only MPI one-sided operations.
        }
        \label{fig:OneSided_Windows}
    \end{center}
\end{figure}

In order to enable remote memory communications during the aforementioned execution phases, we define a multi-window configuration per MPI process (\autoref{fig:OneSided_Windows}):

\begin{itemize}
    \item \emph{``Status'' Window}. Defines the current status for each individual process (e.g., ``\texttt{STATUS\_REDUCE}''). The status is updated remotely after completing a phase.
    \item \emph{``Key-Value'' Window}. This multi-dimensional, dynamic window contains buckets to store the key-value pairs, indexed by the target rank.
    \item \emph{``Combine'' Window}. Designed for the Combine phase, it contains a single-dimension, dynamic window with ordered key-values.
    \item \emph{``Displacement'' Window}. Two additional displacement windows are defined to support the ``Key-Value'' and ``Combine'' windows described\footnote{Attaching new allocations to an MPI dynamic window is not a collective operation. Thus, the MPI standard requires applications to share the \emph{displacement} for the buckets attached to the window by other means.}.
\end{itemize}

When a new key-value pair is found, we use a custom \emph{memory management} to store the correspondent \texttt{<key,value>} tuple. Each key-value pair is mapped inside the current bucket assigned to the target process. We use this approach as a mechanism to transfer information \mbox{concurrently}~\cite{hoefler2009towards}. The target is determined by first generating a 64-bit hash of the key. Thereafter, a mapping to the associated chunk inside the Key-Value window is established. Thus, remote processes can directly reference specific key-values, without affecting the information stored in surrounding buckets. The information is encoded by including a fixed-size header \texttt{h} with the length of the \texttt{key} and \texttt{value} attributes. This fact implies that our implementation supports variable-length \texttt{<key,value>} tuples, of arbitrary $K$ and $V$ bytes, respectively:

\begin{center}
\vspace{0.21cm}
\begin{tikzpicture}[scale=0.6921]
    \draw[solid, fill=white] (0,0) rectangle node{\footnotesize{\texttt{h}}} +(0.81,0.64);
    \draw[solid, fill=white] (0.81,0) rectangle node{\footnotesize{\texttt{key}}} +(2.21,0.64);
    \draw[solid, fill=white] (3.021,0) rectangle node{\footnotesize{\texttt{value}}} +(3.81921,0.64);
    \draw [decorate,decoration={brace,amplitude=4pt},xshift=0,yshift=-2.921] (0.76921,0.0) -- (0.0421,0.0) node [black,midway,yshift=-9.21] {\footnotesize $H_{bytes}$};
    \draw [decorate,decoration={brace,amplitude=4pt},xshift=0,yshift=-2.921] (2.9621,0.0) -- (0.87921,0.0) node [black,midway,yshift=-9.21] {\footnotesize $K_{bytes}$};
    \draw [decorate,decoration={brace,amplitude=4pt},xshift=0,yshift=-2.921] (6.78921,0.0) -- (3.06921,0.0) node [black,midway,yshift=-9.21] {\footnotesize $V_{bytes}$};
\end{tikzpicture}
\vspace{-0.21cm}
\end{center}

\begin{figure}
    \begin{center}
        \includegraphics[width=0.921\columnwidth]{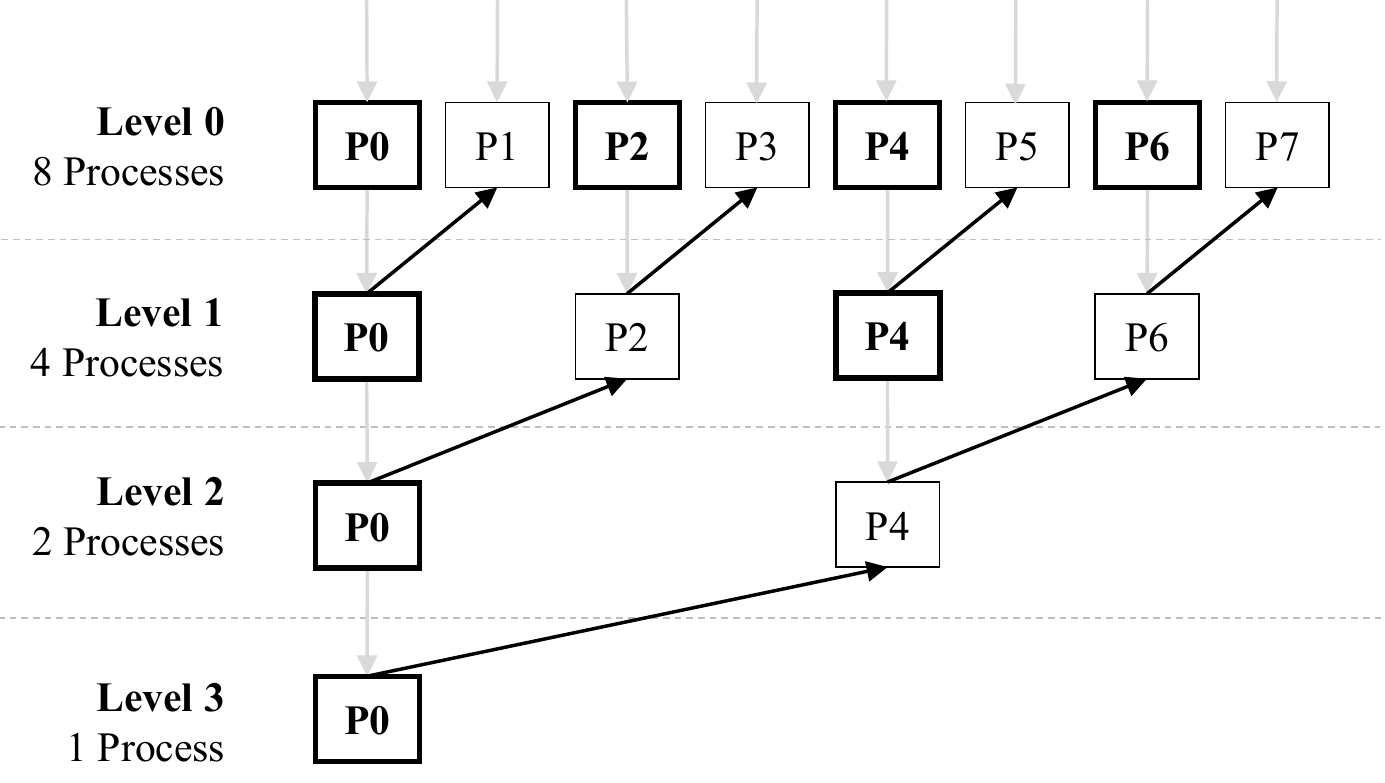}
        \vspace{0.21cm}
        \caption{Example of the tree-based algorithm that generates the final result in Combine, inspired by merge sort~\cite{cole1988parallel, davidson2012efficient}. For illustration purposes, only 8 processes are depicted.}
        \label{fig:Combine}
    \end{center}
\end{figure}


When a process finishes the Map phase, it proceeds to the Reduce phase by collecting groups of key-value pairs assigned to this particular process, from all the other processes. The key-values are retrieved using MPI one-sided operations over the Key-Value window with an \emph{offset}. The offset is estimated using the rank and the specific bucket information defined within the Displacement window. We use the \emph{passive} target synchronization for efficient, decoupled communication~\cite{gropp2014using, gerstenberger2014enabling}. After one group of key-value pairs is retrieved, the process splits the information by interpreting the headers and reducing locally the \texttt{<key,value>} tuples.

In this regard, the Status window is required as a synchronization mechanism to prevent incorrect data accesses to the Key-Value window. The status changes are notified via an atomic \texttt{put} operation. This is accomplished with a combination of \texttt{MPI\_Accumulate} plus \texttt{MPI\_REPLACE} to enforce atomicity~\cite{MPI3standard2015}.
Hence, when a key-value pair is emitted and about to be stored inside Map, we ensure first that the status of the target process that owns the key-value is not already in Reduce. In such case, we avoid to update the bucket, and the ownership of the key-value is transferred\footnote{Despite the ownership change, the key-value will be reduced afterwards during the final Combine phase.}.

\begin{figure}
\vspace{-0.21cm}
\begin{lstlisting}
...
// Create MR object with MR-1S back-end
MapReduce1S *map_reduce =
                       new WordCount();

// Init job with the input settings
map_reduce->Init(filename, win_size,
                 chunk_size, task_size,
                 s_enabled, h_enabled,
                 api, sfactor, sunit);

// Launch execution and output result
map_reduce->Run();
map_reduce->Print();

// Close job and release memory
map_reduce->Finalize();
delete map_reduce;
...
\end{lstlisting}
\captionof{lstlisting}{Source code example in C++ that illustrates how to run a Word-Count job using MapReduce-1S as back-end.\label{exampleCode}}
\end{figure}

Lastly, after the Map and Reduce phases are completed, the Combine phase sets up a tree-based sorting algorithm that fetches the final key-values of each process to generate the result (\autoref{fig:Combine}). We use an algorithm inspired by \emph{merge sort}~\cite{cole1988parallel, davidson2012efficient}. The number of levels in the tree is given by $\lceil\log_2({numProcs})\rceil + 1$. The initial level stores the local key-value pairs in-order. After this first step, the processes retrieve the remote key-values from the previous level using one-sided operations and generate a new level with all the pairs ordered. This task is repeated until one last process generates the final level, which corresponds to the result.

We note that, at this point, we require to enforce synchronization to prevent race conditions over the different levels of the tree. To overcome this limitation with MPI one-sided communication, we use an \emph{exclusive} lock
(i.e., \texttt{MPI\_LOCK\_EXCLUSIVE}) over the Combine window. The lock is acquired by each process during initialization and released after the Combine phase is completed. This guarantees that remote processes are blocked by the MPI implementation until the Combine window is unlocked and the access epoch is completed.

\subsection{Custom MapReduce Framework}

In order to support our experiments, we integrate MapReduce-1S as the back-end of a custom MapReduce framework\footnote{\url{https://github.com/sergiorg-kth/mpi-mapreduce-1s}}. The implementation is written in C/C++ and consists of approximately 1,500 lines of code. 

The framework employs a multi-inheritance mechanism by dividing the responsibilities as a hierarchy of classes:

\begin{itemize}
    \item \textit{Base Class}. Defines the main API to interact with the user, such as initialization or job execution.
    \item \textit{Back-end Class}. Contains the back-end implementation that performs the phases of the algorithm.
    \item \textit{Use-case Class}. Exposes the specific \texttt{Map()} and \texttt{Reduce()} functions required for MapReduce.
\end{itemize}

This set of abstract classes allows applications to easily configure different back-ends over multiple use-cases. Listing~\ref*{exampleCode} provides a source code example in C++ where a Word-Count job is created using MapReduce-1S as back-end. The example first creates the \texttt{WordCount} object, that contains the specific definition of \texttt{Map()} and \texttt{Reduce()}, as well as \texttt{ReduceLocal()}. These functions will be called by MapReduce-1S as necessary. Thereafter, the MapReduce job is initialized by providing several settings, such as the size of each individual task within the Map phase, or the maximum number of bytes that can be transferred simultaneously from remote processes during Reduce and Combine. The execution is then launched and the output result printed.




\subsubsection{Reference MapReduce Implementation}

We also integrate into our custom framework a MapReduce-2S (i.e., MapReduce ``Two-Sided'') implementation based on the work by Hoefler et al.~\cite{hoefler2009towards}. 
In this case, the implementation uses \texttt{MPI\_Scatter} to distribute the tasks using a master-slave approach. The input is read collectively with MPI I/O to optimize the access to storage. During Reduce, \texttt{MPI\_Alltoallv} is used to distribute the variable-length key-value pairs assigned to each process. 
The Combine step follows the same tree-based algorithm of MapReduce-1S, but using point-to-point communication instead. 
The mapping and reduction mechanisms for each key-value pair are also identical. 
This includes the optimizations (e.g., Local Reduce), as well as the custom memory management based on multiple buckets per process.

\section{Experimental Results}
\label{3_Results}

In this section, we estimate the overall performance of MapReduce-1S in comparison with MapReduce-2S. We aim to understand how our approach could be integrated into current and future MapReduce frameworks. After this section, we continue and extend the discussion on the results.

For this purpose, we use \emph{Tegner}, a supercomputer at KTH Royal Institute of Technology with 46 compute nodes. Each node is equipped with dual 12-core Haswell E5-2690v3 processors running at 2.6GHz, and a total of 512GB DRAM. The storage employs a Lustre parallel file system (client v2.5.2) with 165 OST servers. No local storage is provided per node. The OS is CentOS v7.4.1708 with Kernel 3.10.0-693.11.6.el7.x86\_64. The framework is compiled with Intel ICC and Intel MPI, both v18.0.1.

Note that all the figures reflect the standard deviation of the samples as error bars. In addition, we neglect from our results the initialization time, but account for the time required to retrieve the input datasets and bucket allocation. 
Lastly, the terms ``MR-1S'' and ``MR-2S'' are used to refer to MapReduce-1S and MapReduce-2S, respectively.

\input{3_Results_1_fig1}
\begin{figure*}[t]
    \centering
    \begin{subfigure}[t]{0.4921\textwidth}
        \centering
        \begin{tikzpicture}
            \begin{axis}[
                xlabel=MPI Processes / Dataset Size,
                ylabel=Execution Time (s),
                symbolic x coords={8,16,32,64,128,256,512},
                xticklabel style={text width=1.81cm},
                xticklabel style={align=center},
                xticklabels={
                    16\\\vspace{-0.2921cm}\scriptsize{(32GB)},
                    32\\\vspace{-0.2921cm}\scriptsize{(32GB)},
                    64\\\vspace{-0.2921cm}\scriptsize{(32GB)},
                    128\\\vspace{-0.2921cm}\scriptsize{(32GB)},
                    256\\\vspace{-0.2921cm}\scriptsize{(32GB)},
                },
                xtick=data,
                ytick={0,60,120,180,240,300},
                ymin=0,
                ymax=300,
                scaled y ticks = false,
                y tick label style={/pgf/number format/fixed, /pgf/number format/1000 sep = },
                ylabel style={at={(0.021,0.521)}, style={font=\small}},
                xlabel style={at={(0.5,-0.21921)}, style={font=\small}},
                enlarge x limits=0.16,
                grid=major,
                legend style={legend columns=-1,at={(1.0,1.321)},anchor=north east},
                legend style={/tikz/every even column/.append style={column sep=4.21}},
                legend cell align=right,
                legend plot pos=right,
                legend style={draw=none, fill=white, inner xsep=0, inner ysep=0},
                ybar=0pt,
                bar width=6.0921pt,
                width=0.81\textwidth,
                height=3.81cm,
                area legend
            ]
            
            \draw (0.4921cm,0.0cm) |- (0.6421cm,1.9621cm) [solid, draw=black!81] node [right, fill=white, inner xsep=1.21, inner ysep=1.21] {\scriptsize{Multi-Node $\rightarrow$}};
            
            \addplot [fill=black!9] plot[error bars/.cd, y dir=both, y explicit] table [y error minus=min, y error plus=max] {
                x       y               min             max
                16	217.9988852	1.4514784407	1.4514784407
                32	112.2269843	5.1664717645	5.1664717645
                64	57.2565591	2.9305122325	2.9305122325
                128	28.9663957	1.386242308	1.386242308
                256	14.9687475	0.5166781246	0.5166781246
            }; \addlegendentry{MR-1S}
            
            \addplot [fill=black!9, postaction={pattern=crosshatch dots}] plot[error bars/.cd, y dir=both, y explicit] table [y error minus=min, y error plus=max] {
                x       y               min             max
                16	239.7501989	2.9556406197	2.9556406197
                32	116.6329056	0.8352501594	0.8352501594
                64	58.6094484444	0.6011316213	0.6011316213
                128	30.2183363	1.3260175873	1.3260175873
                256	16.8817517	1.2685987086	1.2685987086
            }; \addlegendentry{MR-1S (FT)}
            \end{axis}
        \end{tikzpicture}
        \caption{\textbf{Strong Scaling Evaluation} with Checkpoints} \label{fig:benchmarks_checkpoint_strong}
    \end{subfigure}
    \hfill
    \begin{subfigure}[t]{0.4921\textwidth}
        \centering
        \begin{tikzpicture}
            \begin{axis}[
                xlabel=MPI Processes / Dataset Size,
                ylabel=Execution Time (s),
                symbolic x coords={8,16,32,64,128,256,512},
                xticklabel style={text width=1.81cm},
                xticklabel style={align=center},
                xticklabels={
                    16\\\vspace{-0.2921cm}\scriptsize{(16GB)},
                    32\\\vspace{-0.2921cm}\scriptsize{(32GB)},
                    64\\\vspace{-0.2921cm}\scriptsize{(64GB)},
                    128\\\vspace{-0.2921cm}\scriptsize{(128GB)},
                    256\\\vspace{-0.2921cm}\scriptsize{(256GB)},
                },
                xtick=data,
                ytick={0,40,80,120,160,200},
                ymin=0,
                ymax=200,
                scaled y ticks = false,
                y tick label style={/pgf/number format/fixed, /pgf/number format/1000 sep = },
                ylabel style={at={(0.021,0.521)}, style={font=\small}},
                xlabel style={at={(0.5,-0.21921)}, style={font=\small}},
                enlarge x limits=0.16,
                grid=major,
                legend style={legend columns=-1,at={(1.0,1.321)},anchor=north east},
                legend style={/tikz/every even column/.append style={column sep=4.21}},
                legend cell align=right,
                legend plot pos=right,
                legend style={draw=none, fill=white, inner xsep=0, inner ysep=0},
                ybar=0pt,
                bar width=6.0921pt,
                width=0.81\textwidth,
                height=3.81cm,
                area legend
            ]
            
            \draw (0.4921cm,0.0cm) |- (0.6421cm,1.9621cm) [solid, draw=black!81] node [right, fill=white, inner xsep=1.21, inner ysep=1.21] {\scriptsize{Multi-Node $\rightarrow$}};
            
            \addplot [fill=black!9] plot[error bars/.cd, y dir=both, y explicit] table [y error minus=min, y error plus=max] {
                x       y               min             max
                16	95.1999163	1.1361324601	1.1361324601
                32	115.452564	7.9707692454	7.9707692454
                64	120.3687229	3.4975033758	3.4975033758
                128	121.1075426	7.3056705154	7.3056705154
                256	126.1597624	21.600641757	21.600641757
            }; \addlegendentry{MR-1S}
            
            \addplot [fill=black!9, postaction={pattern=crosshatch dots}] plot[error bars/.cd, y dir=both, y explicit] table [y error minus=min, y error plus=max] {
                x       y               min             max
                16	97.4508711111	0.6611352092	0.6611352092
                32	116.6266922	0.8228860992	0.8228860992
                64	124.8985224444	1.1109740887	1.1109740887
                128	129.6164115	7.1804039211	7.1804039211
                256	131.4528382222	4.6073878437	4.6073878437
            }; \addlegendentry{MR-1S (FT)}
            \end{axis}
        \end{tikzpicture}
        \caption{\textbf{Weak Scaling Evaluation} with Checkpoints} \label{fig:benchmarks_checkpoint_weak}
    \end{subfigure}
    \caption{Strong / Weak scaling performance with variable-size datasets from PUMA-Wikipedia using MapReduce-1S on Tegner, under balanced workload. The results illustrate the original version of this implementation and the modified version with checkpoint support through MPI storage windows.\label{fig:benchmarks_checkpoint_strong_weak}}
\end{figure*}
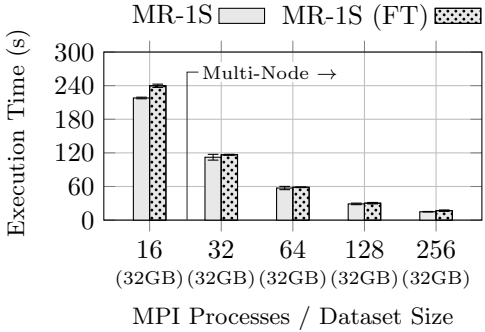
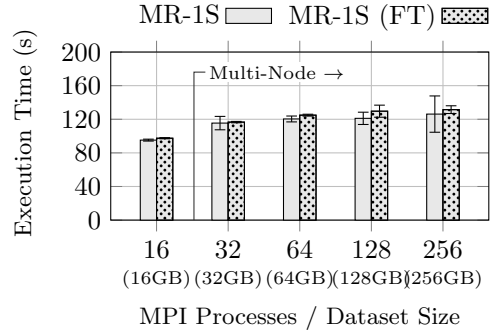
\input{4_Discussion_fig1}

\subsection{Performance Evaluation}

We evaluate the scalability of MapReduce-1S using Word-Count~\cite{ranger2007evaluating, ahmad2012puma}, a technique that has major relevance in Big Data analytics. For instance, Word-Count has been proposed in the past for understanding the quality of articles on Wikipedia~\cite{blumenstock2008size}. The basic principle of Word-Count is to compute the occurrences of individual words over large collections of documents. Here, the Map phase emits \texttt{<word,1>} key-value pairs, where \texttt{word} represents the key and \texttt{1} the occurrence found. 
The Reduce phase aggregates the occurrences for a given \texttt{word} to generate its final \texttt{<word,count>}. Finally, the Combine phase aggregates the key-values to produce the result.

For our evaluations, we use a large dataset from the Purdue MapReduce Benchmarks Suite (PUMA)~\cite{ahmad2012puma}. This suite emerges as an on-going effort to provide rigorous benchmarks for MapReduce frameworks. 
In particular, we use the \emph{Dataset3} from the PUMA-Wikipedia datasets\footnote{\url{https://engineering.purdue.edu/~puma/datasets.htm}}, that contains approximately 300GB of data divided into multiple files. These files include articles, user discussions, and other metadata originally from Wikipedia. We pre-process the files off-line to generate unified, large input datasets for concise results. Thus, allowing us to have fine-grained control over the workload assigned per process to evaluate both balanced and unbalanced workloads\footnote{Unbalanced workloads are simulated by computing the same task multiple times, but reading the input only once. Otherwise, we would mostly account for the time required to retrieve the data from storage.}. 
Finally, we use a task size of 64MB, a limit of 1MB per one-sided operation, and an initial bucket size of 64MB per process. The input files are created with a stripe size of 1MB and maximum stripe count (165). The settings are determined empirically for each implementation prior to running our experiments.

Using a fixed-size input dataset, we observe that MapReduce-1S scales ideally when duplicating the number of active processes, but does not provide significant performance advantages on large process counts. Figures~\ref*{fig:benchmarks_strong_weak}\subref*{fig:benchmarks_strong_b1} and \ref*{fig:benchmarks_strong_weak}\subref*{fig:benchmarks_strong_b0} illustrate the performance of MapReduce-2S and MapReduce-1S by varying the number of MPI processes on Tegner for a fixed-size input dataset (strong scaling), using balanced and unbalanced workloads, respectively. The process count varies from 16 (1 node) up to 256 (11 nodes). We use a reference input dataset from PUMA-Wikipedia with 32GB of data. From Figure~\ref*{fig:benchmarks_strong_weak}\subref*{fig:benchmarks_strong_b1}, we determine that, if the workload is ideally balanced, MapReduce-1S provides approximately 4.8\% improvement on average over MapReduce-2S for lower process counts (up to 64 processes). If the workload is unbalanced, Figure~\ref*{fig:benchmarks_strong_weak}\subref*{fig:benchmarks_strong_b0} shows that the average improvement is approximately 20.4\% on average compared to MapReduce-2S. However, the performance of MapReduce-1S is affected on large process counts due to the limited workload per process (e.g., 0.1GB on the last test). In such situations, the use of collective communication and collective I/O clearly results in better performance.

By increasing the size of the input datasets and, consequently, the workload per process, we confirm that MapReduce-1S provides performance advantages on unbalanced workloads. Figures~\ref*{fig:benchmarks_strong_weak}\subref*{fig:benchmarks_weak_b1} and \ref*{fig:benchmarks_strong_weak}\subref*{fig:benchmarks_weak_b0} illustrate the performance of MapReduce-2S and MapReduce-1S by varying the number of MPI processes on Tegner and maintaining the workload per process (weak scaling), using balanced and unbalanced workloads, respectively. Once again, the process count varies from 16 (1 node) up to 256 (11 nodes). We use the reference input datasets from PUMA-Wikipedia, with 16GB up to 256GB of data (i.e., 1GB per process). When the workload is ideally balanced, we conclude that MapReduce-1S provides equivalent performance compared to MapReduce-2S, as illustrated in Figure~\ref*{fig:benchmarks_strong_weak}\subref*{fig:benchmarks_weak_b1}. The average execution time is 111.3 seconds for MapReduce-2S, and 111.8 seconds for MapReduce-1S (0.5\% difference). Nonetheless, when the workload per process is unbalanced, we observe evident performance benefits on all the experiments. The average execution times are 649.9 seconds for MapReduce-2S, and 530.8 seconds for MapReduce-1S. The improvement is 23.1\% on average, with a peak of 33.9\%. 

\section{Discussion}
\label{4_Discussion}

We further extend the discussion concerning the results given in the previous section.



\subsubsection*{Considerations for Map and Reduce phases}

The experimental results have illustrated that overlapping the Map, Reduce, and Combine phases in MapReduce-1S, can provide benefits when the workload per process becomes unbalanced. We have additionally observed that the use of MPI one-sided communication and individual non-blocking I/O incurs in a performance penalty on large process counts if the workload per process is limited or balanced. In such cases, the use of collective communication and I/O still provides performance advantages in comparison. However, we must note that the benefits of MapReduce-1S directly depend on the particular use-case. In Word-Count, for example, the execution is largely dominated by the Map phase, with lightweight Reduce and Combine phases (i.e., execution mostly depends on the time required to retrieve the input). 
We expect to analyze additional use-cases in future work, with the aim of understanding the implications of our approach compared to other MapReduce frameworks.

\input{4_Discussion_fig3}

\subsubsection*{Opportunities for fault-tolerant MapReduce}

The use of MPI one-sided communication inside MapReduce-1S provides us with an opportunity to integrate the MPI storage windows concept~\cite{rivas2017mpi}. This project is an on-going effort that proposes the use of MPI one-sided communication and MPI windows as unique interface to program data movement among memory and storage subsystems. The approach transparently integrates storage into the memory management of HPC applications, requiring only subtle source code modifications on already existing applications that use the MPI one-sided communication model. Hence, we could easily define a novel fault-tolerant MapReduce-1S implementation that establishes a mapping to storage per window (i.e., transparent checkpoint). \autoref*{fig:benchmarks_checkpoint_strong_weak} illustrate the strong / weak scaling performance of MapReduce-1S on Tegner, under balanced workload. We introduce support for MPI storage windows and extend MapReduce-1S to perform a window synchronization point\footnote{In MPI storage windows, applications can guarantee data consistency with the storage layer through \texttt{MPI\_Win\_sync}. We use this function as a mechanism to ensure that the latest window changes are flushed to storage.} after each Map task, as well as after the Reduce phase is completed. From these figures, we determine that the checkpoint overhead is only 4.8\% on average. The reason for this optimal result is due to the fact that, with MPI storage windows, data transferring from / to storage is overlapped with computations. The synchronization points are only required to ensure consistency.

\subsubsection*{Memory requirements per dataset}

One of the main limiting factors of MapReduce in the context of HPC is the high memory requirements of this model~\cite{gao2017mimir}. In the case of MapReduce-1S, the need for pre-allocating multiple window buckets to enable one-sided operations might also become a constraint. Nonetheless, we employ several optimizations that aim to reduce the memory footprint. 
Figure~\ref*{fig:discussion_memory}a shows the peak memory consumption per node on Tegner using the MapReduce-2S and MapReduce-1S implementations during the weak scaling evaluation. The workload per process is 1GB (24GB per node). From this figure, we determine that both implementations reflect similar memory requirements between 10.4--13.7GB. The peak consumption is observed during Combine at the end of the execution (Figure~\ref*{fig:discussion_memory}b).

\subsubsection*{Importance of the MPI implementation}

Even though we use the passive target synchronization of MPI one-sided communication, clear communication patterns are observed when we analyze the execution timeline of our experiments. The behaviour is similar to the \emph{active} target synchronization, which is close to traditional point-to-point communication. Thus, limiting the performance benefits of our approach. In order to partially reduce these constraints, we included redundant lock / unlock operations for each window after all the Map and Reduce tasks. Figure~\ref*{fig:discussion_timeline} illustrates an example execution timeline of MapReduce-1S before and after this change. Despite the performance improving approximately 5\% on average, evident communication patterns still exist. This effect was observed with recent versions of both Intel MPI and OpenMPI implementations. We plan to investigate the behaviour of other MPI implementations in the future.

\section{Related Work}
\label{5_RelatedWork}

Since its inception, the MapReduce programming model has been widely adopted by the scientific community~\cite{dean2008mapreduce, ekanayake2008mapreduce}. For instance, Chu et al.~\cite{chu2007map} propose MapReduce to parallelize a variety of machine learning algorithms, such as Locally Weighted Linear Regression (LWLR). 

In the context of HPC, state-of-the-art MPI functionality is employed to take advantage of the high-performance network and storage subsystems~\cite{plimpton2011mapreduce}. Guo et al.~\cite{guo2015fault} propose a fault-tolerant MapReduce implementation that uses fine-grained progress tracking to establish locally consistent states for failure recovery. Recently, Gao et al.~\cite{gao2017mimir} provide an efficient MapReduce library designed to reduce the overall memory footprint on current and future supercomputers. While the goal of the aforementioned publications is mostly to bridge the gap between data analytics and scientific computing, we consider our approach complimentary. Hence, the integration of some of the developments described in this paper could provide further value to their proposals.

Lastly, we must note that the work by Hisham et al.~\cite{mohamed2013mro} share some similarities. Here, the authors propose a MapReduce framework specifically designed to overlap the Map and Reduce phases to decrease the constraints on imbalanced workloads. Their proposal is based on running the Map and Reduce phases in parallel, and exchanging partial intermediate results between each phase using MPI. In our case, we take a different direction and propose to decouple these phases completely with the integration of MPI one-sided communication and non-blocking I/O instead.

\section{Conclusion}
\label{6_Conclusion}

With the emergence of machine learning and data-centric applications on HPC, MapReduce has become one of the preferred programming models to hide the complexity of process and data parallelism~\cite{dean2008mapreduce, plimpton2011mapreduce}. In this paper, we have presented a decoupled strategy for MapReduce frameworks based on the integration of MPI one-sided communication and non-blocking I/O operations~\cite{gropp2014using,thakur1999data}.

Preliminary results have demonstrated that, while our approach does not provide significant performance benefits on large-process counts and balanced workloads per process, it does feature performance advantages by overlapping the Map and Reduce phases of MapReduce if the workload per process becomes unbalanced.


As future work, we plan to analyze the implications of a job stealing mechanism based on the use of atomic MPI one-sided operations. In addition, we plan to investigate the integration of fault-tolerance support on MapReduce-1S through the concept of MPI storage windows~\cite{rivas2017mpi}.


 \section*{Acknowledgements}
The experimental results were performed on resources provided by the Swedish National Infrastructure for Computing (SNIC) at PDC Centre for High Performance Computing (PDC-HPC).

The work was funded by the European Commission through the SAGE project (Grant agreement no. 671500 / http://www.sagestorage.eu). 


\bibliography{main}

\begin{thebibliography}{10}
\expandafter\ifx\csname url\endcsname\relax
  \def\url#1{\texttt{#1}}\fi
\expandafter\ifx\csname urlprefix\endcsname\relax\def\urlprefix{URL }\fi
\expandafter\ifx\csname href\endcsname\relax
  \def\href#1#2{#2} \def\path#1{#1}\fi

\bibitem{sage2016whitepaper}
{SAGE Consortium}, {Data Storage for Extreme Scale - The SAGE Project White
  Paper}, \url{http://bit.ly/2eSYPRH}, [On-Line] (2016).

\bibitem{gao2017mimir}
T.~Gao, Y.~Guo, B.~Zhang, P.~Cicotti, Y.~Lu, P.~Balaji, M.~Taufer, {Mimir:
  Memory-efficient and scalable MapReduce for large supercomputing systems},
  in: Parallel and Distributed Processing Symposium (IPDPS), 2017 IEEE
  International, IEEE, 2017, pp. 1098--1108.

\bibitem{szegedy2015going}
C.~Szegedy, W.~Liu, Y.~Jia, P.~Sermanet, S.~Reed, D.~Anguelov, D.~Erhan,
  V.~Vanhoucke, A.~Rabinovich, Going deeper with convolutions, in: Proceedings
  of the IEEE conference on computer vision and pattern recognition, 2015, pp.
  1--9.

\bibitem{witten2016data}
I.~H. Witten, E.~Frank, M.~A. Hall, C.~J. Pal, Data Mining: Practical machine
  learning tools and techniques, Morgan Kaufmann, 2016.

\bibitem{dean2008mapreduce}
J.~Dean, S.~Ghemawat, {MapReduce: Simplified data processing on large
  clusters}, Communications of the ACM 51~(1) (2008) 107--113.

\bibitem{ekanayake2008mapreduce}
J.~Ekanayake, S.~Pallickara, G.~Fox, {MapReduce} for data intensive scientific
  analyses, in: eScience, 2008. eScience'08. IEEE Fourth International
  Conference on, IEEE, 2008, pp. 277--284.

\bibitem{hoefler2009towards}
T.~Hoefler, A.~Lumsdaine, J.~Dongarra, Towards efficient {MapReduce} using
  {MPI}, in: European Parallel Virtual Machine/Message Passing Interface
  Users’ Group Meeting, Springer, 2009, pp. 240--249.

\bibitem{tu2008scalable}
T.~Tu, C.~A. Rendleman, D.~W. Borhani, R.~O. Dror, J.~Gullingsrud, M.~O.
  Jensen, J.~L. Klepeis, P.~Maragakis, P.~Miller, K.~A. Stafford, et~al., A
  scalable parallel framework for analyzing terascale molecular dynamics
  simulation trajectories, in: High Performance Computing, Networking, Storage
  and Analysis, 2008. SC 2008. International Conference for, IEEE, 2008, pp.
  1--12.

\bibitem{shvachko2010hadoop}
K.~Shvachko, H.~Kuang, S.~Radia, R.~Chansler, {The Hadoop Distributed File
  System}, in: Mass storage systems and technologies (MSST), 2010 IEEE 26th
  symposium on, Ieee, 2010, pp. 1--10.

\bibitem{reed2015exascale}
D.~A. Reed, J.~Dongarra, Exascale computing and big data, Communications of the
  ACM 58~(7) (2015) 56--68.

\bibitem{plimpton2011mapreduce}
S.~J. Plimpton, K.~D. Devine, {MapReduce in MPI for Large-scale Graph
  Algorithms}, Parallel Computing 37~(9) (2011) 610--632.

\bibitem{guo2015fault}
Y.~Guo, W.~Bland, P.~Balaji, X.~Zhou, Fault tolerant {MapReduce-MPI} for {HPC}
  clusters, in: Proceedings of the International Conference for High
  Performance Computing, Networking, Storage and Analysis, ACM, 2015, p.~34.

\bibitem{dongarra2011international}
J.~Dongarra, P.~Beckman, T.~Moore, P.~Aerts, G.~Aloisio, J.-C. Andre,
  D.~Barkai, J.-Y. Berthou, T.~Boku, B.~Braunschweig, et~al., {The
  International Exascale Software Project Roadmap}, The International Journal
  of High-Performance Computing Applications 25~(1) (2011) 3--60.

\bibitem{peng2017preparing}
I.~B. Peng, R.~Gioiosa, G.~Kestor, E.~Laure, S.~Markidis, Preparing hpc
  applications for the exascale era: A decoupling strategy, in: Parallel
  Processing (ICPP), 2017 46th International Conference on, IEEE, 2017, pp.
  1--10.

\bibitem{thakur1999data}
R.~Thakur, W.~Gropp, E.~Lusk, Data sieving and collective {I/O} in {ROMIO}, in:
  Frontiers of Massively Parallel Computation, 1999. Frontiers' 99. The Seventh
  Symposium on the, IEEE, 1999, pp. 182--189.

\bibitem{gropp2014using}
W.~Gropp, T.~Hoefler, R.~Thakur, E.~Lusk, {Using advanced {MPI}: Modern
  features of the message-passing interface}, MIT Press, 2014.

\bibitem{ahmad2012puma}
F.~Ahmad, S.~Lee, M.~Thottethodi, T.~Vijaykumar, {PUMA: Purdue MapReduce
  Benchmarks Suite}.

\bibitem{rivas2017mpi}
S.~Rivas-Gomez, R.~Gioiosa, I.~B. Peng, G.~Kestor, S.~Narasimhamurthy,
  E.~Laure, S.~Markidis, {MPI Windows on Storage for HPC Applications}, in:
  Proceedings of the 24th European MPI Users' Group Meeting, ACM, 2017, p.~15.

\bibitem{chu2007map}
C.-T. Chu, S.~K. Kim, Y.-A. Lin, Y.~Yu, G.~Bradski, K.~Olukotun, A.~Y. Ng,
  {Map-Reduce for machine learning on multicore}, in: Advances in neural
  information processing systems, 2007, pp. 281--288.

\bibitem{lu2011can}
X.~Lu, B.~Wang, L.~Zha, Z.~Xu, {Can MPI benefit Hadoop and MapReduce
  applications?}, in: Parallel Processing Workshops (ICPPW), 2011 40th
  International Conference on, IEEE, 2011, pp. 371--379.

\bibitem{gerstenberger2014enabling}
R.~Gerstenberger, M.~Besta, T.~Hoefler, Enabling highly-scalable remote memory
  access programming with {MPI-3} one sided, Scientific Programming 22~(2)
  (2014) 75--91.

\bibitem{cole1988parallel}
R.~Cole, Parallel merge sort, SIAM Journal on Computing 17~(4) (1988) 770--785.

\bibitem{davidson2012efficient}
A.~Davidson, D.~Tarjan, M.~Garland, J.~D. Owens, {Efficient parallel merge sort
  for fixed and variable length keys}, in: Innovative Parallel Computing
  (InPar), 2012, IEEE, 2012, pp. 1--9.

\bibitem{MPI3standard2015}
{Message Passing Interface Forum}, {{MPI}: A Message-Passing Interface
  Standard}, Vol. 3.1, 2015,
  \url{http://mpi-forum.org/docs/mpi-3.1/mpi31-report.pdf}. Accessed:
  2018-02-21.

\bibitem{ranger2007evaluating}
C.~Ranger, R.~Raghuraman, A.~Penmetsa, G.~Bradski, C.~Kozyrakis, {Evaluating
  MapReduce for multi-core and multiprocessor systems}, in: High Performance
  Computer Architecture, 2007. HPCA 2007. IEEE 13th International Symposium on,
  Ieee, 2007, pp. 13--24.

\bibitem{blumenstock2008size}
J.~E. Blumenstock, {Size matters: Word Count as a measure of quality on
  Wikipedia}, in: Proceedings of the 17th international conference on World
  Wide Web, ACM, 2008, pp. 1095--1096.

\bibitem{mohamed2013mro}
H.~Mohamed, S.~Marchand-Maillet, {MRO-MPI: MapReduce overlapping using MPI and
  an optimized data exchange policy}, Parallel Computing 39~(12) (2013)
  851--866.

\end{thebibliography}

\end{document}